\documentclass[apj]{emulateapj}
\journalinfo{The Astrophysical Journal, 690: L97-L100, 2009 January 10}

\usepackage{epsfig,times}

\newcommand{\msun}{M_{\odot}} 
\newcommand{\Rs}{R_{\mathrm{S}}}
\newcommand{\nh}{N_{\mathrm{H}}}
\newcommand{\ginj}{\Gamma_{\rm inj}}
\newcommand{\gel}{\Gamma_{\rm e}}
\newcommand{\etab}{\eta_{\mathrm{B}}}
\newcommand{\Tbb}{T_{\mathrm{BB}}}
\newcommand{\Te}{T_{\mathrm{e}}}

\newcommand{\Linj}{L_{\mathrm{inj}}}
\newcommand{\Ldisk}{L_{\mathrm{disk}}}
\newcommand{\UB}{U_{\mathrm{B}}}
\newcommand{\Urad}{U_{\mathrm{rad}}}

\newcommand{\me}{m_{\rm e}} 
 
\newcommand{\sigmaSB}{\sigma_{\mathrm{SB}}} 
 
\newcommand{\taup}{\tau_{\mathrm{p}}} 
\newcommand{\Ne}{N_{\mathrm{e}}} 

\newcommand{\alphax}{\alpha} 
 
\newcommand{\dotgamma}{\dot{\gamma}}

\shorttitle{Spectral states of accreting black holes}
\shortauthors{Poutanen \& Vurm}

\begin{document}

\title{On the origin of spectral states in accreting black holes} 

\author{Juri Poutanen\altaffilmark{1} and Indrek Vurm\altaffilmark{1,2}}

\altaffiltext{1}{Astronomy Division, Department of Physical Sciences, P.O.Box 3000, 
90014 University of Oulu, Finland; juri.poutanen@oulu.fi,  indrek.vurm@oulu.fi}
\altaffiltext{2}{Tartu Observatory, 61602 T\~{o}ravere, Tartumaa, Estonia; vurm@ut.ee}

\submitted{Received 2008 July 19; accepted 2008 November 20; published 2008 December 15}

\begin{abstract} 
The origin of dramatically different electron distributions responsible for Comptonization in black hole X-ray binaries (BHBs) in their various states is discussed. We solve the coupled kinetic equations for photons and electrons without approximations on the relevant cross-sections accounting for Compton scattering, synchrotron radiation, and Coulomb collisions. In the absence of external soft photons,  the electrons are efficiently thermalized  by synchrotron self-absorption and Coulomb scattering  even for pure nonthermal electron injection. The resulting quasi-thermal synchrotron self-Compton spectra have very  stable slopes and electron temperatures similar to the hard states of BHBs.  The hard spectral slopes observed in the X-rays, the cutoff at 100 keV, and the MeV tail  together require low magnetic fields ruling out the magnetic dissipation mechanism. The motion of the accretion disk toward the black hole results in larger Compton cooling and lower equilibrium electron temperature.  Our self-consistent simulations show that in this case both electron and photon distributions attain a power-law-dominated shape similar to what is observed in the soft state. The electron distribution in the Cyg X-1 soft state  might require a strong magnetic field, being consistent with the magnetically dominated  corona. 
\end{abstract}


\section{Introduction}

The physical processes giving rise to the X-ray/gamma-ray emission of accreting black-hole binaries (BHBs) have been a matter of debates over the last four decades. The hard-state spectra, showing a  strong cut-off  around 100 keV, are well described by thermal Comptonization \citep[e.g.][]{P98,ZG04}, while a weak MeV tail  requires the presence of nonthermal particles \citep{McConnell94,Ling97}. The origin of seed soft photons for Comptonization is, however, much less clear. An apparent correlation between the spectral slope and the amount of Compton reflection \citep{ZLS99} argues in favor of the accretion disk, while the observed optical/X-ray correlation \citep{MIC82,Kanbach01} leans toward  the synchrotron hypothesis \citep[e.g.][]{FGM82,WZ01}. Interesting questions are then: what stabilizes the X-ray spectral slope at $\alphax\sim 0.6$--0.8, and what  fixes the temperature of the emitting plasma at $k\Te\sim$50--100 keV \citep{ZJP97,P98,ZG04}? Do the feedback from the cool accretion disk and the thermostatic properties of electron-positron pairs \citep{HM93,HMG94,SPS95,MBP01}  play a role here? Or does the cooling by synchrotron radiation \citep{NY95} act as a stabilizer?  

In the soft state, BHB spectra are dominated by thermal disk emission of temperature $k\Tbb\sim$0.4--1.5 keV. At higher energies the spectrum is power-law-like and shows no signatures of the cut-off  \citep{Grove98} extending possibly up to 10 MeV \citep{McConnell02}. This emission is well described by Comptonization in almost purely nonthermal plasmas \citep{PC98,G99,coppi99, ZGP01,ZG04}.  We can then ask why the electrons are nearly thermal in the hard state, and what causes such a dramatic change in the electron distribution when transition to the soft state happens. \citet{PC98} proposed that the two states are distinguished by the way the energy is supplied to the electrons: by thermal heating, dominating during the hard state, and by nonthermal acceleration, operating in the soft state. However, their treatment of Coulomb collisions \citep[using {\sc eqpair} code by][]{coppi92,coppi99} was approximate, and they have neglected the  effect of synchrotron boiler, involving the emission and absorption of synchrotron photons, which can act as an efficient particle thermalizer \citep{GGS88}.

\citet{GHS98} studied for the first time the combined effect of the synchrotron boiler and Compton cooling on the electron distribution and photon spectra (but neglected Coulomb scattering). They considered a two-phase corona model \citep{HM93,HMG94,SPS95}, where half of the high-energy radiation was assumed to be reprocessed by the disk to soft photons. As the actual geometry of the emitting region is not known, we start from pure synchrotron self-Compton models (i.e. with no external soft photons) and compute self-consistently the electron (positron) and photon distributions.  We then investigate  how the additional soft photons (e.g., associated with the inner radius of  the  cool accretion disk) affect the equilibrium distributions and compare the results of simulations with the data on Cyg X-1. The preliminary results of this study were presented by \citet{VP08a}.

\section{Model setup} 

We consider a  black hole of mass 10 $\msun$, typical for stellar-mass BHBs. We assume that the inner  accretion flow is hot and almost spherical, corresponding to the advection-dominated \citep{NY95,ACK95,Esin97} or to the recently discovered luminous hot accretion flow solutions \citep{yuan03,yuan_aaz04}.  One expects that most of the gravitational energy release happens within about $R=10\Rs=3\times 10^7$ cm (where $\Rs=2GM/c^2$ is the Schwarzschild radius) from the black hole, and we thus fix the size of the active region at this value in most of the simulations. The released energy needs to be transferred to electrons via, e.g., Coulomb collisions with hot protons, collective plasma effects, magnetic reconnection, or shocks.   We assume that the energy transfer to the electrons is given by a power-law-injection function $d\Ne/(dt\ d\gamma)$$\propto$$\gamma^{-\ginj}$ extending in the Lorentz factor from  $\gamma=1$ to $10^3$. To keep the Thomson optical depth  of the electrons associated with protons    $\taup$ constant,\footnote{The total optical depth might be larger due to the produced pairs, but for parameters considered here, the amount of pairs is negligible.}  the same number of electrons from the equilibrium  distribution is removed from the system.   In this case, the net power is $\Linj$=$\frac{4\pi}{3} R^3 \dot{\Ne} ( \langle \gamma \rangle_{\rm inj} -\langle \gamma \rangle_{\rm eq} )  \me c^2$, where $\langle \gamma \rangle_{\rm inj}$  and $\langle \gamma \rangle_{\rm eq}$ are the mean Lorentz factors of the injection function and of the equilibrium distribution, respectively, and $\dot{\Ne}$ is uniquely determined by the model parameters and $\langle\gamma\rangle_{\rm eq}$.

The injected electrons  are cooled by synchrotron emission and Compton scattering at timescales much shorter than the accretion time. The synchrotron radiation is strongly self-absorbed up to hundreds of harmonics, and therefore the cooling depends strongly on the high-energy tail of the electron distribution \citep[see e.g.][]{WZ01}. The importance of synchrotron processes is determined by the ratio $\etab $=$\UB R^2 c/\Linj$, where $ \UB$=$B^2/(8\pi)$ is the magnetic energy density and $\Linj\approx\frac{4\pi}{3}R^2 c \Urad$ (so that  $\etab$$\approx$$3/4\pi$$\approx$0.25 corresponds to an equipartition of the magnetic and radiation energy densities, $\UB$=$\Urad$). The seed photons for Compton upscattering can be provided by the synchrotron as well as by the external sources, the cool accretion disk being the most natural one.  The external soft photons are modeled as a blackbody of temperature $\Tbb$ determined from the Stefan-Boltzmann law $\Ldisk$=$4\pi R^2 \sigmaSB \Tbb^4$. The cooling by external photons depends on the ratio $f$=$\Ldisk/\Linj$.  The total escaping photon luminosity is $L$=$\Ldisk+ \Linj= (1+f) \Linj$.

To model self-consistently particle and photon distributions, we solve numerically a set of coupled, time-dependent kinetic equations for photons, electrons, and positrons describing Compton scattering, cyclo-synchrotron emission and absorption, electron-electron Coulomb (M{\o}ller) scattering as well as  pair production and annihilation. Our calculations are done in a simple one-zone geometry with a tangled magnetic field and isotropic particle and photon distributions. The escape probability formalism is used to simulate photon escape from the region. The detailed description of  the code and its extensive testing are given in \citet{VP08b}.

\begin{figure*}
\centerline{\epsfig{file= 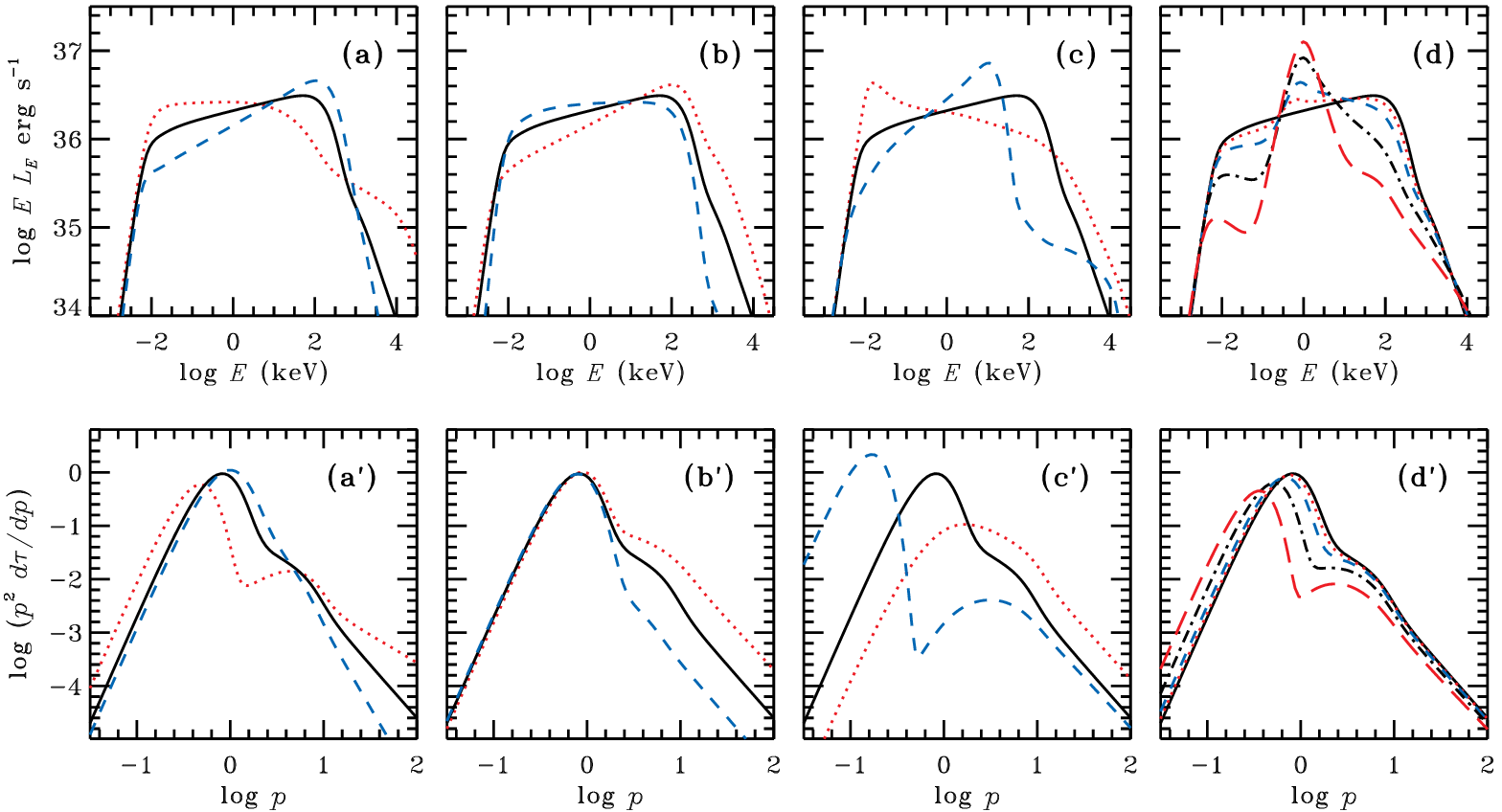,width=16.5cm}} 
\caption{Equilibrium photon spectra (upper panels) and electron distributions (lower panels) $p^2 d\tau/dp$ (i.e., momentum per log $p$, where $\tau$ is the Thomson optical depth and $p=\sqrt{\gamma^2-1}$ is the dimensionless electron momentum).   
The results for the fiducial parameter set $L=\Linj=10^{37}$ erg s$^{-1}$, $\ginj=3$, $\taup=1.5$, $\etab=1$,  $f=0$, 
are shown by solid curves in all panels.
(a-a') Dependence on the electron injection slopes: $\ginj=2$ (dotted curve), 3 (solid) and 4 (dashed).  
(b-b') Dependence on magnetization $\etab=0.1$ (dotted), 1 (solid),  10 (dashed). 
(c-c') Dependence on the optical depth  $\taup=0.15$ (dotted), $1.5$  (solid), 15 (dashed). 
(d-d') Dependence on the ratio of the external disk photons to the injected power $f=0, 0.1, 0.3, 1, 3$ 
(solid, dotted, dashed, dot-dashed, long-dashed curves, respectively) for constant total luminosity $L$.
The electron temperatures and photon spectral indices  are given in Table \ref{tbl-1}.   
}	
\label{fig:all}
\end{figure*}

\section{Synchrotron self-Compton models}

We first assume that the cool disk  is sufficiently far away and does not supply any seed soft photons to the inner hot flow. 
Thus we consider pure synchrotron self-Compton (SSC) models ($f$=0). We choose the fiducial parameter set  $L=\Linj=10^{37}$ erg s$^{-1}$ and $\taup$=1.5 (typical for the hard state  of BHBs, \citealt{ZJP97}), $R=3\times10^7$ cm, $\ginj=3$ (ad hoc),  and $\etab$=1. The equilibrium electron distribution consists of a Maxwellian part with $k\Te$=66 keV and a power-law-like tail with the slope  modified by cooling $\gel=\ginj +1=4$ (where $d\Ne/d\gamma\propto \gamma^{-\gel}$; see the solid curve in Fig.~\ref{fig:all}a'). The synchrotron emission  is strongly self-absorbed with only  the nonthermal tail above $\gamma\gtrsim$20 contributing to emission above the self-absorption energy at $\gtrsim$10 eV.  As the amount of seed soft (synchrotron) photons is low, the  Comptonization spectrum (produced predominantly by the thermal electron population) is hard with  the photon energy index $\alphax\approx0.9$ and a cut-off at $\sim$100 keV, which is similar to the hard state of BHBs. A tail produced by single-Compton scattering off the power-law electron tail is clearly visible above MeV. 

\begin{table}
\begin{center}
\caption{Results of simulations \label{tbl-1}}
\begin{tabular}{cccr}
\tableline\tableline
 \multicolumn{2}{c}{P\tablenotemark{a}}  & $\alphax$\tablenotemark{b} & $\Te$\tablenotemark{c} \\
      &  &        & (keV)     \\
\tableline
  \multicolumn{2}{c}{Fiducial\tablenotemark{d}}    &  0.89 &  66  \\
\tableline  
$\ginj$ & 2&  1.07 &  27  \\
           & 4  &  0.73 &  90 \\
\tableline 
 $\etab$& 0.1  &  0.75 &  77  \\
 	& 10  &  0.98 &  64 \\
\tableline
 $\taup$ & 0.15 &  1.12 &  160  \\
 	    &15     &  0.58 &  4 \\
\tableline
 $f$        & 0.1 &  0.97 &  61 \\
              & 0.3 &  1.13 &  49 \\
 	    & 1    &  1.61 &  31 \\
  	    & 3    & 2.46 &  16 \\
\tableline
\end{tabular}
\tablenotetext{a}{Varying parameter and its value.}
\tablenotetext{b}{Photon spectral index in the 2--10 keV range. }
\tablenotetext{c}{Temperature of the Maxwellian part of the electron distribution.}
\tablenotetext{d}{Fiducial set of parameters $L=10^{37}$ erg s$^{-1}$, $R=3\times 10^7$ cm, 
$f=0$, $\ginj=3$, $\taup=1.5$, $\etab=1$.}
\end{center}
\end{table}

Variation of the slope of the injected electrons leads to a large change in the tail of the electron distribution and dramatic difference in the synchrotron emission. A soft electron injection with $\ginj\gtrsim4$ leads to efficient thermalization and a small amount of soft photons resulting in rather hard  radiation spectra, with the photon energy index $\alphax\lesssim0.7$  (see dashed curves in Fig.~\ref{fig:all}a). A hard injection gives more power to the nonthermal tail and more seed photons for Comptonization \citep[see also][]{GHS98,WZ01}, which causes a drop in the electron temperature (see the dotted curves in Fig.~\ref{fig:all}a).  A strong 'bump' also develops in the tail of the electron distribution at $\gamma\sim$3.   The synchrotron emission produced by these electrons is still strongly self-absorbed, while the energy losses and gains stay close to each other for an extended energy interval \citep{KGS06}. In this regime, the ratio of synchrotron heating and cooling rates for a power-law distribution of relativistic electrons is $\dotgamma_{\rm h}/\dotgamma_{\rm c} \approx 5/(\gel+2)$.\footnote{This expression can be derived  by employing the delta-function approximation for synchrotron emissivity to calculate the source function and using it in the expression for heating by self-absorption \cite[see e.g.][]{GGS88, KGS06}.} Observe that for $\gel=$3 (i.e. for $\ginj$=2) the heating and cooling rates are balanced, however, such an equilibrium is unstable \citep{Rees67}. The Comptonized spectrum for hard injection $\ginj=$2 (see Fig.~\ref{fig:all}a) is much softer than the hard-state spectra of BHBs, even without any contribution to the cooling from the disk, strongly constraining the electron injection mechanism in BHBs. 
 
The efficiency of synchrotron cooling depends on the magnetic field strength parameterized here via magnetization $\etab$.  At small $\etab$ (see  Fig.~\ref{fig:all}b'), synchrotron is inefficient and cooling is dominated by thermal Comptonization. A higher normalization of the power-law part of the equilibrium electron distribution leads to a stronger MeV tail. For $\etab\lesssim$1 (and $\ginj>3$), the thermal Comptonization spectrum is very stable with $\alphax\sim$0.7--0.9. At large $\etab>1$, the synchrotron thermalization operates more efficiently and the thermal part of the distribution persists to higher energies. The increasing $B$ field compensates for the decrease in the power-law tail leading to a higher synchrotron emission,  which results in softening of the Comptonized spectrum.

Consider now variations of $\taup$ for the fixed $\Linj$. At high $\taup$, the equilibrium electron temperature drops, leaving fewer energetic electrons for synchrotron emission and, therefore, reducing the number of seed photons for Comptonization (Fig.~\ref{fig:all}c). This, in turn, results in the harder photon spectra produced by saturated Comptonization (by thermal electrons) and a weak high-energy tail (produced by nonthermal electrons), very similar to the ultrasoft spectra of BHBs \citep[see figs. 8 and 9 in][]{ZG04}. At smaller $\taup$,  the higher electron temperature leads to a stronger synchrotron cooling and to a lower Comptonized luminosity  and, therefore,  softer Comptonized spectra.

Let us now apply the developed model to the hard state of Cyg X-1. The MeV tail observed there with $\alpha_{\rm MeV}\approx2$ \citep{McConnell02} constrains the injection slope to be $\ginj <2\alpha_{\rm MeV}=4$.  Then the  hard X-ray spectra with $\alphax\lesssim$0.7 and a high-energy cutoff at $\sim$100 keV (see Fig. \ref{fig:st_tra}) require $\taup\sim 1$ and low $\etab \lesssim$0.1 \citep[see also][]{WZ01,McConnell02}. Any additional soft photons from the disk will make the spectrum softer, reducing $\etab$ even more. The low magnetic field rules out magnetic reconnection as the energy dissipation mechanism. This also implies that electrons cannot be thermalized by the synchrotron self-absorption. On the other hand, if the size of the active region is $R$$\sim$$60\Rs$, Coulomb  scattering becomes important (as its influence grows linearly with size for constant $L$, see, e.g., \citealt{coppi99,sve99}), and it can thermalize electrons at a rather high temperature of  $k\Te$$\sim$100 keV as observed in Cyg X-1 \citep{G97,P98}.

We reiterate that the whole spectrum here is produced by the SSC mechanism.\footnote{ADAF-based models also consider SSC as the main cooling mechanism; see \citet{NMQ98} for the review.}  Its  thermostatic properties fix the electron temperature at 50--100 keV (for $\taup\sim1$) and stabilize the spectral slope at $\alphax$$\sim$0.7--0.9. The feedback from the disk \citep{HM93,HMG94,SPS95,MBP01} does not seem to be needed.


\section{Spectral transitions and the role of  disk photons} 

\begin{figure}
\centerline{\epsfig{file= 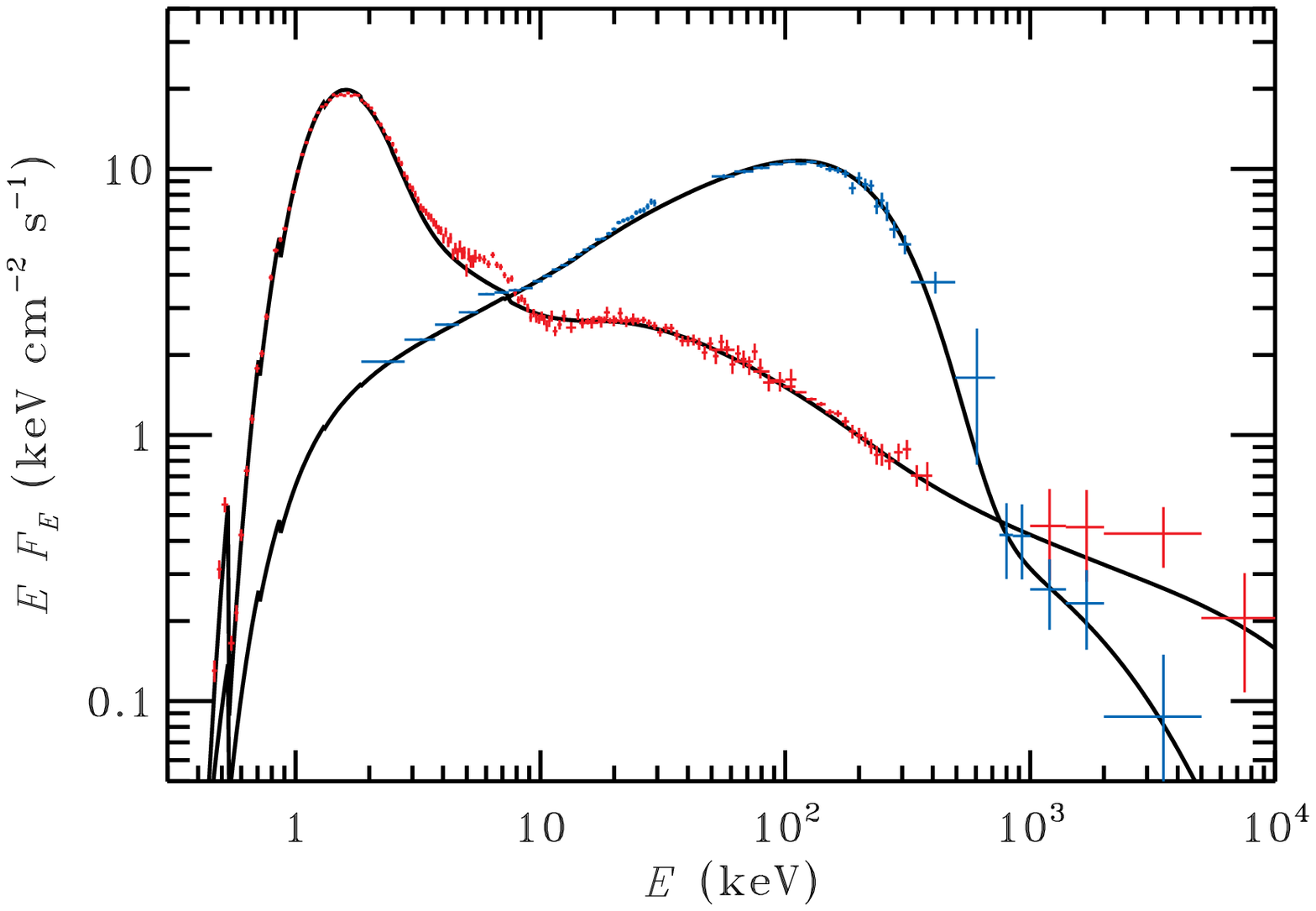,width=8.5cm} } 
\caption{Spectral states of Cyg X-1. Crosses show the unfolded spectral data presented by \citet{ZPP02}. 
The model spectra with interstellar absorption (described by the column density $\nh$) and Compton reflection (described by the solid angle $\Omega$ and ionization parameter $\xi$, \citealt{MZ95}) taken into account are shown by the solid curves. The parameters for the hard-state model are: $L=2.7\times 10^{37}$ erg s$^{-1}$, $R=1.8\times10^8$ cm, $f=0,$ $\ginj=3.8$, $\taup=2.5$, $\etab=0.083$, $\nh=3\times10^{21}$ cm$^{-2}$, $\Omega/(2\pi)=0.2$, $\xi=0$. 
The soft state can be described by $L=4.85\times10^{37}$ erg s$^{-1}$, $R=3.85\times10^7$ cm, 
$f=2.13$, $\ginj=2.2$, $\taup=0.3$, $\etab=1.5$, $\nh=5\times10^{21}$ cm$^{-2}$, $\Omega/(2\pi)=0.7$, $\xi=100$. 
}
\label{fig:st_tra}
\end{figure}

The spectral transitions observed in BHBs are most probably accompanied by a change in the geometry of the accretion disk. The cool outer disk moves toward the central black hole causing an increasing flux of the soft photons to the central hot flow  \citep{Esin97,PKR97,PC98}, which we simulate here by increasing $f$ (see Fig.~\ref{fig:all}d). Higher soft photon flux leads to faster Compton cooling and lower equilibrium electron temperature, making the nonthermal part more pronounced. The resulting photon distribution changes from the hard, thermal Comptonization  dominated, spectrum to the one dominated by the disk blackbody, with a nonthermal tail extending to tens of MeV, which becomes harder at higher $f$. The spectral changes triggered by varying $f$ are similar to the one observed in Cyg X-1 (see Fig. \ref{fig:st_tra}). A detailed comparison with Cyg X-1 spectra shows, however, that other parameters change too. 
 
Compared to the hard state, the soft state corresponds to a higher total luminosity. The  MeV tail is harder $\alpha_{\rm MeV} \approx1.6$ \citep{McConnell02}, and therefore  $\ginj <3.2$. If the tail of the black-body at 3--10 keV (see Fig. \ref{fig:st_tra}) is produced in the same emission region, it requires a rather hot thermal population of electrons, which needs high $\etab$ for the synchrotron thermalization  to operate (because Coulomb thermalization is not efficient  under the conditions of strong Compton cooling).  This would be consistent with the magnetically dominated emission region. Alternatively, there may be additional heating mechanisms operating. Also the tail might be a result of Comptonization in the hot ionized skin of the disk, not directly related to the emission we discuss here, but this interpretation might not be easily reconciled with the fact that the disk is stable, while the tail varies \citep{chur01}. While the dramatic changes in the electron and photon distributions between the states are mainly caused by variations of the disk luminosity, it is obvious that other parameters do change during the transition. We stress that none of the presented models requires any additional thermal heating, which is different from the models of  \citet{PC98} and \citet{G99}. 


\section{Conclusions}

The hard state of BHBs can well be described by the quasi-thermal SSC mechanism. The feedback from the cool disk is not needed to stabilize the spectral slope and the electron temperature. Electrons can be injected to the active region with the power-law spectrum, but Coulomb scattering and synchrotron self-absorption thermalize them efficiently. This reduces the need  for mysterious 'thermal heating' that was invoked previously to explain thermal Comptonization spectra of BHBs. The MeV tail together with the hard X-ray spectra of BHBs with photon indices $\alphax\lesssim$0.7 and a cutoff at 100 keV require rather low magnetization $\etab<0.1$ and a large size of $R\gtrsim 60\Rs$.   In that case, magnetic reconnection can be ruled out as a source of energy. 
 
At high optical depth of the emitting region $\taup\gtrsim10$, in the absence of disk radiation, the spectrum is close to saturated Comptonization, peaking at a few keV. This Wien-type spectrum might be associated with the ultrasoft  state of BHBs. At low $\taup$, the electrons are hotter and the spectra are softer due to the efficient synchrotron cooling.

A behavior similar to what is observed during the spectral state transitions in BHBs can be reproduced by varying the ratio of injected soft luminosity and the power dissipated in the hot phase, which could be caused by varying the radius of the inner cool disk. The increasing Compton cooling causes dramatic changes in the electron distribution from almost purely thermal to nearly nonthermal.  The photon distribution also changes from quasi-thermal SSC  to the nonthermal Comptonization of the disk photons. In the soft state of Cyg X-1,  a strong magnetic field can thermalize electrons at sufficiently high temperature, which is consistent with a magnetically dominated corona being responsible for the high-energy emission.

\acknowledgments
 
We are grateful to the anonymous referee and Andrzej Zdziarski for a number of useful comments. 
This study was supported by the CIMO grant TM-06-4630 and the Academy of Finland grants 110792  and 122055.



\end{document}